\documentclass[amsmath,aps,prb,superscriptaddress,showpacs,amssymb,secnumarabic,eqsecnum,twocolumn,nofootinbib,floatfix]{revtex4-1}
\usepackage{graphicx}
\usepackage{pstricks}
\usepackage{natbib,hyperref,url}
\usepackage{version}

\newrgbcolor{blanco}{255 255 255}
\newcommand{\ba}{\begin{eqnarray}}
\newcommand{\ea}{\end{eqnarray}}
\def\be{\begin{equation}}
\def\ee{\end{equation}}

\def\vk{\mathbf{k}}

\begin{document}
\title{Generalized susceptibilities and Landau parameters for anisotropic Fermi liquids}
\author{Pablo Rodr\'\i guez-Ponte}
\affiliation{Instituto de F\'\i sica de La Plata and Departamento de F\'isica, Universidad Nacional de La Plata, C.C. 67, 1900 La Plata, Argentina.}
\author{Nicol\'as Grandi}
\affiliation{Instituto de F\'\i sica de La Plata and Departamento de F\'isica, Universidad Nacional de La Plata, C.C. 67, 1900 La Plata, Argentina.}
\affiliation{Abdus Salam International Centre for Theoretical Physics, Associate Scheme, \\ Strada Costiera 11, 34151, Trieste, Italy.}
\author{Daniel C. Cabra}
\affiliation{Instituto de F\'\i sica de La Plata and Departamento de F\'isica, Universidad Nacional de La Plata, C.C. 67, 1900 La Plata, Argentina.}
\affiliation{Abdus Salam International Centre for Theoretical Physics, Associate Scheme, \\ Strada Costiera 11, 34151, Trieste, Italy.}
\begin{abstract}
We study Fermi liquids with a Fermi surface that lacks continuous rotational invariance, and in the presence of an arbitrary quartic interaction. We obtain the expressions of the generalized static susceptibilities that measure the linear response of a generic order parameter to a perturbation of the Hamiltonian. We apply our formulae to the spin and charge susceptibilities.  Based on the resulting expressions, we make a proposal for the definition of the Landau parameters in non-isotropic Fermi liquid.
\end{abstract}
\maketitle
\section{Introduction}

Fermi-liquid theory\cite{Legget,Baym,nozieres} is a standard paradigm to describe many-body fermionic systems. The basic assumption is that the weakly excited states of a Fermi liquid are in one to one correspondence to the weakly excited states of a Fermi gas. The states of the Fermi gas can be described by means of elementary excitations with momenta close to the Fermi surface. By adiabatically switching on the two-particle interaction, we can realize the correspondence to the Fermi liquid states by introducing the concept of quasi-particles. These quasi-particles are assumed to have long lifetimes, and their distribution function in momentum space shows a step at finite momentum that defines the Fermi surface.

In Landau's original description, the Fermi surface has continuous rotational invariance. Thus, the interaction strength between quasiparticles of different momenta depends only on the angle spanned by the momenta. This implies that it can be decomposed in a basis of functions of this single variable. When the basis is that of Legendre polynomials, the coefficients of such expansion define the so-called ``Landau parameters''. This parameters are useful in writing the thermodynamic responses (such as spin- or charge-susceptibility, specific heat, etc.) in terms of the ones corresponding to the non-interacting case. Moreover, at zero temperature \cite{nos5} the Landau parameters play a central role in the Pomeranchuk criteria to diagnose Fermi liquid instabilities that lead to continuous phase transitions. 

In practice, however, there are several interesting cases where the Fermi surface lacks of continuous rotational invariance, the simplest example being that of a fermionic system defined on an arbitrary lattice. Another example are the symmetry-broken phases that arise as a consequence of electron-electron interactions\cite{Metzner0,Metzner01,Gros,Wu,Wuotro,Bonfim,Quintanilla3,Jaku} which include nematic phases, a possibility that has attracted much attention recently \cite{nematics,nematic1,*nematic1a,*nematic1b,*nematic1c,Metzner_PRL,nematic2,Hankevych,Metzner1,Wu1,nematic_MF,YK,Lawler,Lawlerb,Kee1,nematic3,Metzner2,Metzner2b,Quintanilla2}. In particular, the $Sr_3Ru_2O_7$ compound at low temperatures, exhibits an intermediate phase\cite{grigeraotrosa,*grigeraotrosb,*grigeraotrosc,grigera2011,grigera2012} which has been shown to be compatible with a nematic Fermi liquid\cite{yamase2005,yamase,yamase2010,yamase2007,*yamaserestob,*yamaserestoc}.

When the Fermi surface is not rotational invariant, the detection of Pomeranchuk instabilities becomes more subtle\cite{nos5}. The same is true for the calculation of thermodynamic responses. An open question is whether it is possible to define generalized Landau parameters playing a similar role than those in the isotropic case. Previous calculations of susceptibilities and proposals for the Landau  parameters corresponding to anisotropic Fermi liquids are found in Ref. [\onlinecite{DasSarmaanisotropic}], where the case of dipolar interaction is analyzed in detail under a perturbative approach. In \onlinecite{Fuseya}, Fuseya {\em et. al.} proposed definitions for the first Landau parameters corresponding to the non-isotropic case for the two dimensional Hubbard model. Appart from these two papers, we are not aware of further studies of this issue and, in particular, no systematic analysis can be found in the literature.

In the present paper we address this issue in a systematic way and derive the equations that allow to compute all linear response functions in a generic Fermi liquid. Our analysis leads to a natural definition of the Landau parameters that extend the rotationally invariant expression.
We study the linear response functions of an anisotropic Fermi liquid in the presence of generic interactions by proposing a general decomposition for the interaction function. By comparing the results with those of the isotropic case we propose definitions for the generalized Landau parameters. We verify that such proposal plays in the anisotropic Pomeranchuk criterion at zero temperature a similar role than the usual Landau parameter in an isotropic setup. 

The paper is organized as follows: in section \ref{sec:model} the decomposition of the interaction function as well as the resulting renormalized dispersion relation are discussed. The generalized susceptibilities are explicitly calculated in section \ref{sec:generalizedsusc} and the generalization of Landau parameters is proposed. Its role on Pomeranchuk instabilities is compared with that in the isotropic case. Some application examples are presented in section \ref{sec:interactions}. 

\section{Quasiparticle interaction and Mean Field theory.}
\label{sec:model}
Our aim is to calculate thermodynamic response functions for anisotropic Fermi Liquids. Thus, we work with a generic Hamiltonian containing a four fermion 
interaction term. It reads
\be
\begin{split}
\hat{H} &= \sum_{\vk,\alpha}  \epsilon_o^\alpha(\vk) \hat{c}^\dagger_{\alpha,\vk} \hat{c}_{\alpha, \vk}  \\
&+ \sum_{\vk,\vk',\textbf{q}} \sum_{\alpha,\beta,\gamma, \delta} f^{\alpha\beta\gamma \delta}(\vk,\vk',\textbf{q})  \hat{c}^\dagger_{\alpha, \vk+\textbf{q}} \hat{c}^\dagger_{\gamma, \vk'-\textbf{q}} \hat{c}_{\delta, \vk'} \hat{c}_{\beta, \vk}, 
\end{split}
\label{hamil}
\ee
where the operator $\hat{c}^\dagger_{\alpha\vk}$ ($\hat{c}_{\alpha\vk}$) 
creates (destroys) a fermion with momenta $\vk$ and spin  $\alpha$. Here $\epsilon_o(\vk)$ is the bare dispersion relation. Finally, $f^{\alpha\beta\gamma\delta}\!(\mathbf{k},\mathbf{k'},\mathbf{q})$ is a tensor characterizing the interaction. 

Given the form of the Hamiltonian (\ref{hamil}), the renormalized dispersion relation is obtained by minimizing the free energy  within a mean field approximation (details are given in Appendix \ref{sec:meanfield}). It reads
\ba
\epsilon^\alpha(\vk)&=& \epsilon_o^\alpha(\vk) + \sum_{\beta,\vk'} f^{\alpha \beta}(\vk,\vk') n_{\beta}(\vk'). 
\label{selfcon1}
\\
n_{\alpha}(\vk)&=&F\left[\frac{\epsilon^\alpha(\mathbf{k})}{k_BT}\right]
\label{selfcon2}
\ea
Where the ``interaction function'' $f^{\alpha\beta}(\mathbf{k},\mathbf{k}')$ is defined as $f^{\alpha\beta}(\mathbf{k},\mathbf{k}')=f^{\alpha\alpha\beta\beta}(\mathbf{k},\mathbf{k}',\mathbf{0})$. Notice that these are coupled equations that must be solved self consistently.

When working with an isotropic Fermi liquid,  in which no underlying lattice is considered, it is usual to decompose the interaction function in spin symmetric and antisymmetric parts, as
\be \label{standardecomposition}
f^{\alpha\beta}\!(\mathbf{k},\mathbf{k'})=f^{(s)}(\vk,\vk') + \alpha\beta f^{(a)}(\vk,\vk'). 
\ee
Being the Fermi surface defined by the relation $\epsilon_o(\vk)=0$, it has continuous rotational symmetry when $\epsilon_o(\vk)=\epsilon_o(|\vk|)$. As the interaction includes only low energy excitations, it can be written just as a function of the angle between $\vk$ and $\vk'$, 
\be
f^{(s,a)}\!(\mathbf{k},\mathbf{k'})=f^{(s,a)}(\cos{\theta_{\vk\vk'}}).
\ee
Furthermore, it is customary to expand the interaction function in Legendre polynomials (or cosine functions for the 2D case) and define the Landau parameters as the dimensionless coefficients of such expansion. For example, for three-dimensional systems the usual definition is as follows\cite{Legget}
\be 
\label{standardlandauparameters}
F^{(s,a)}_l=\frac{2l+1}{2} N(0)  \int f^{(s,a)}(\cos{\theta_{\vk\vk'}})P_l{ (\cos{\theta_{\vk\vk'}}) } d\theta_{\vk\vk'},
\ee
where $N(0)$ is the density of states at the Fermi surface and $P_l$ are the Legendre polynomials. These parameters are useful to diagnose zero-temperature instabilities on the Fermi liquid phase according to the Pomeranchuk criteria\cite{Pomeranchuk}
\be
1+F_l^{(s,a)}>0\,.
\label{lachota}
\ee
Moreover they allow us to write the thermodynamic responses in terms of those obtained on the non-interacting limit. For example, the spin and charge susceptibilities are given by \cite{nozieres}
\be \label{isotropicexpressions}
\begin{split}
\chi_{spin}=\frac{dM}{dh}&=\chi_{spin}^0 (1+F_0^a)^{-1}
\\
\chi_{charge}=\frac{dN}{d\mu}&=\chi_{charge}^0 (1+F_0^s)^{-1},
\end{split}
\ee
where $M$ is the magnetization, $N$ the particle number, $h$ the magnetic field, and $\mu$ the chemical potential. Here $\chi_{spin}^0$ and $\chi_{charge}^0$ are the susceptibilities corresponding the non-interacting case.

~

However, when analyzing more realistic models suitable to describe real compounds, the Fermi surface has only the discrete rotational invariance associated with the symmetries of the underlying lattice. In this context it is not possible to write the interaction function in terms of the single variable $\theta_{\vk\vk'}$. Instead, we work with the decomposition
\be 
\label{interac}
f^{\alpha\beta}(\vk,\vk')=-\sum_{i,j=1}^N U_{ij}d^\alpha_i(\vk)d^\beta_j(\vk')
\ee
in terms of a basis of functions $\{d_i^\alpha({\bf k})\}_{i\in\mathbb{N}}$. 
We assume that the decomposition has a finite number of terms $N$, which holds for most of the interactions studied in the literature. 
Here and in what follows, Latin indexes run from $1$ to $\infty$, {\em i.e.} they go through all the basis elements $i,j,k,l\in \mathbb{N}$, while we keep in mind that the matrix $U_{ij}$ has a finite number of non-vanishing components.
As $f^{\alpha\beta}(\vk,\vk')=f^{\beta\alpha}(\vk',\vk)$, the interaction coefficients in expression \eqref{interac} are symmetric in their indexes, so  $U_{ij}=U_{ji}$.

~

Inserting the expression for the decomposed form of the interaction function, eq.(\ref{interac}) onto the self consistent equations (\ref{selfcon1}) and (\ref{selfcon2}), we obtain 
\be
\epsilon^\alpha(\vk)=\epsilon_o^\alpha(\vk)
-\sum_{{i}{j}}^N U_{{i}{j}}\eta_{{i}} d_{{j}}^\alpha(\vk).  \label{reldisperrenor} 
\ee
Thus, the dispersion relation is written in terms of the bare dispersion relation $\epsilon_o^\alpha(\vk)$ and the parameters 
$\eta_{{j}}$ defined as
\be
\label{orderparameter} 
\eta_{{j}}=\sum_{\vk,\alpha} \, d_{{j}}^\alpha(\vk) F\left[\frac{\epsilon^\alpha(\vk)}{k_BT}\right],  
\ee
where $F[x]=1/(\exp(x) +1)$ is the Fermi-Dirac distribution, $T$ is the temperature and $k_B$ the Boltzmann constant. Notice that, since the Fermi-Dirac distribution is evaluated on the renormalized dispersion $\epsilon^\alpha(\vk)$, eqs. (\ref{reldisperrenor} - \ref{orderparameter}) must be solved in a self-consistent way.

It is clear from the expression \eqref{reldisperrenor} that the change in the dispersion relation due to interactions has the symmetries dictated by the later. In other words, the parameters $\eta_i,\eta_j$ corresponding to non-vanishing $U_{ij}$ can be understood as order parameters measuring how much has the interaction broken the symmetries of the free system. 

In what follows, we find convenient to reabsorb the temperature in the dispersion relation, so our formula reads
\be
\tilde\epsilon^\alpha(\vk)=\tilde\epsilon_o^\alpha(\vk)
-\sum_{{i}{j}}^N \tilde U_{{i}{j}}\eta_{{i}} d_{{j}}^\alpha(\vk)\,,  \label{reldisperrenor2} 
\ee
where $\tilde \cdots = \cdots/k_BT$.

We are interested in the linear response of the order parameters \eqref{orderparameter} under a change in some of the external control parameters in the partition function, as we explain in the next section.

\section{Generalized susceptibilities.}
\label{sec:generalizedsusc}
The standard computation of susceptibilities involve taking derivatives of order parameters with respect to some control parameter. In this manner, one can obtain the spin susceptibility by taking the derivative of the magnetization $M=\sum_{\vk,\alpha} \alpha n_\alpha(\vk)$ with respect to the magnetic field $h$, or the charge susceptibility by considering the derivative of the number of particles  $N=\sum_{\vk,\alpha} n_\alpha(\vk)$ with respect to the filling $\mu$ of the system. Generalizing, we take the derivative of the order parameter $\eta_j$, with respect to an arbitrary control parameter $a_i$, to obtain the ``generalized susceptibility''
\be 
\label{ec:derivada1}
\chi_{ij}\equiv\frac{d\eta_j}{da_i}.  
\ee
Introducing the expression of the order parameter, eq.(\ref{orderparameter}), we get
\be
\chi_{ij}=\sum_{\vk,\alpha}\,  d_j^\alpha(\vk) F'[\tilde \epsilon^\alpha(\vk)]\frac{d \tilde \epsilon^\alpha (\vk)}{d a_i} ,
\label{ec:derivada2}
\ee
where $F'[x]$ is the derivative of the Fermi distribution with respect to the argument. From equation \eqref{reldisperrenor} we see the dependence of $\tilde \epsilon^\alpha(k)$ on $a_i$. By taking the derivative we write
\be
\frac{d\tilde\epsilon^\alpha(\vk)}{da_i}= \frac {d\tilde \epsilon^\alpha_o(\vk)}{da_i} - \sum_{{ k}{ l}}^N
\left(
\frac{d \eta_{ k}}{d a_{i}}  \tilde U_{{k}{l}}
d_{ l}^\alpha(\vk)  
+
\eta_{ k} 
\frac{d\tilde U_{{k}{l}}}{da_i}
d_{ l}^\alpha(\vk)  
\right) .
\ee
Replacing it into the above equation for the susceptibilities (eq. \eqref{ec:derivada2}), we get
\be
\begin{split}
\chi_{ij} =& \sum_{\vk,\alpha} \, d_j^\alpha(k) F'[\tilde \epsilon^\alpha(k)] \\
\times &\left( \frac {d\tilde \epsilon^\alpha_o(\vk)}{da_i}
- \sum_{{ k}{ l}}
\left(
\frac{d \eta_{ k}}{d a_{i}}  \tilde U_{{k}{l}}
d_{ l}^\alpha(\vk)  
+
\eta_{ k} 
\frac{d\tilde U_{{k}{l}}}{da_i}
d_{ l}^\alpha(\vk)  
\right) 
\right).
\end{split}
\ee
This relation can be simplified by introducing the notation \cite{nos5} 
\be
\label{notation}
\langle \phi|\psi\rangle= \sum_{\vk,\alpha} F'[\epsilon^\alpha(\vk)] \phi^\alpha(\vk) \psi^\alpha(\vk).
\ee
In this manner we obtain 
\be
\chi_{ij}\!= \!  \left\langle \frac {d\tilde \epsilon_o}{da_i}\Big{|}d_j  \right\rangle -\sum_{ {k}{l}} 
\left(\!
\chi_{i{k}}  U_{{k}{l}} \langle d_{{l}} | d_j \rangle 
+
\eta_{{k}} \frac{d\tilde U_{{k}{l}}}{da_i} \langle d_{{l}} | d_j \rangle 
\!\right).
\label{aca}
\ee
This can be solved formally giving the expression for the generalized susceptibilities as
\be 
\label{derivadadeletata}
\boxed{~
\chi_{i{j}}=
\sum_{ r} 
\left(\left\langle \frac {d\tilde \epsilon_o}{da_i}\Big{|}d_{ r} \right\rangle 
-\sum_{ {k}{l}}
\eta_{{k}} \frac{d\tilde U_{{k}{l}}}{da_i} \langle d_{{l}} | d_{ r} \rangle \right)
\Delta_{{r}{j}},  
\phantom{\!\!\!\!\!\frac{\frac{\frac{1^1}2}2}2}
}
\ee
where the matrix $\Delta_{{j}{k}}$ is defined as the inverse matrix
\be
\Delta^{-1}_{\ {j}{k}}=
\delta_{{k}{j}}+
\sum_{{l}}   U_{{k}{l}} \langle d_{{l}} | d_{{j}} \rangle\,.
\label{Delta}
\ee
The expression (\ref{derivadadeletata}) is the main result of this paper. From it, we can calculate the different thermodynamic responses by choosing appropriately the control parameter $a_i$ and the order parameter $\eta_j$. 

An immediate observation is that all the susceptibilities diverge at the points of parameter space at which the inverse \eqref{Delta} does not exists,
\be
\left|
\delta_{{k}{j}}+
\sum_{{l}}   U_{{k}{l}} \langle d_{{l}} | d_{{j}} \rangle 
\right|=0.
\label{det}
\ee
In other words the vanishing of the above determinant signals discontinuous phase transitions. This complements the analysis of Ref.[\onlinecite{nos5}] in which the vanishing of a related expression was shown to be the smoking gun of a continuous (second order) phase transition.

\subsection{Perturbations in the dispersion relation: spin and charge susceptibilities}

As a first example, let us focus in a perturbation on the Hamiltonian proportional to the occupation numbers. So we replace $\hat{H}$ by $\hat{H} +\Delta \hat{H}$ where $\Delta \hat{H}$ is a generic perturbation written as
\be
\Delta \hat{H}=\sum_i a_i \sum_{\vk,\alpha}\,d_i^\alpha(\vk) 
\hat{n}_{\alpha}\!(\vk).
\label{Hext}
\ee
Such perturbation on the Hamiltonian can be understood as a perturbation in the bare dispersion relation $\tilde \epsilon_o^\alpha(\vk)$ which is replaced by 
\be
\tilde \epsilon_o^\alpha(\vk,a_i) =
\tilde\epsilon_o^\alpha(\vk)+\frac 1{k_BT}\sum_i a_i d_i^\alpha(\vk). 
\ee
Then we get for the derivatives on equation (\ref{derivadadeletata}) 
\be
\frac{d \tilde\epsilon_o^\alpha}{da_i}=\frac 1{k_BT}  d_i^\alpha(\vk), 
\qquad \qquad
\frac{dU_{jk}}{da_i}=0,
\ee
implying that the generalized susceptibilities eq.(\ref{derivadadeletata}) in this case read
\be 
\label{derivadadelett}
\chi_{i{j}}=\frac 1{k_BT}  
\sum_{ k} 
\left\langle d_i |d_k\right\rangle 
\Delta_{kj}.  
\ee
This expression depends explicitly on $d_i^\alpha(\vk)$ and $d_j^\alpha(\vk)$, being $d_i^\alpha(\vk)$ the basis function entering into the perturbation, and $d_j^\alpha(\vk)$ the one entering into the order parameter whose response we want to compute. But it also depends on all other basis functions entering into the decomposition of the interaction $d_k^\alpha(\vk)$.

~

We are now able to compute the spin and charge susceptibilities. For the spin susceptibility, we need to calculate the response on the magnetization when the system is exposed to an external magnetic field. In this case we write the perturbation on the Hamiltonian as a Zeeman term
\be
\Delta \hat{H}= h\sum_{\vk,\alpha}\,\alpha 
\hat{n}_{\alpha}\!(\vk),
\label{zeeman}
\ee
which is a particular case of \eqref{Hext} where we take the basis function entering into the perturbation as \break 
$d_a^\alpha(\vk)=\alpha$ with coefficient $a_a=h$. We write the magnetization as 
\be
M\equiv \eta_a=\sum_{\vk,\alpha}\, \alpha F[\epsilon^\alpha(\vk)].
\ee 
Thus, the basis function entering into the order parameter is again $d_a^\alpha(\vk) =\alpha$. The spin susceptibility is defined as 
\be
\chi_{spin}= \frac{dM}{dh}=\frac{d\eta_a}{da_a},
\ee
and using \eqref{derivadadelett} we obtain
\be 
\label{spinsusc}
\chi_{spin}=\frac 1{k_BT}  \sum_{j} \langle  d_a| d_j\rangle 
\Delta_{ja}.
\ee

Similarly, for the charge susceptibility we perturb the Hamiltonian with a chemical potential
\be
\Delta \hat{H}=\mu \sum_{\vk,\alpha}\,  
\hat{n}_{\alpha}\!(\vk)
\ee
which implies that the basis function entering into the perturbation is $d_s^\alpha(\vk) =1$, with coefficient $a_s=\mu$. Then we evaluate the response in the particle number
\be
N\equiv \eta_s = \sum_{\vk,\alpha} \,F[\epsilon^\alpha(\vk)].
\ee
The basis function entering into the order parameter is also $d^\alpha_s(\vk)=1$, in consequence the charge susceptibility obtained from \eqref{derivadadelett} reads
\be 
\chi_{charge}= \frac{dN}{d\mu}=\frac{d\eta_s}{da_s},
\ee
obtaining
\be \label{chargesusc}
\chi_{charge}=\frac 1{k_BT}  \sum_{j} \langle  d_s| d_j\rangle 
\Delta_{js}.
\ee

We can also get the cross-susceptibility, which measures the response on the magnetization under a change in the chemical potential (or conversely the response on the number of particles under a change on the magnetic field). It is straightforward to calculate it identifying the basis function entering into the perturbation with $d_s^\alpha(\vk)=1$ and the one entering into the order parameter with $d_a^\alpha(\vk)=\alpha$ (or vice-versa). We obtain
\be
\chi_{cross}=\frac 1{k_BT}  \sum_{j} \langle  d_s| d_j\rangle 
\Delta_{ja}.
\ee
Then, we see that the cross susceptibility depends on $d_s^\alpha(\vk)$ and $d_a^\alpha(\vk)$ as well as the rest of the basis functions entering in the in the interaction function

\subsection{Landau parameters}
In order to define Landau parameters for the anisotropic case, we first notice that, when considering the non-interacting limit ($U_{ij}=0$ for all $i,j$), then from \eqref{Delta} $\Delta_{ij}=\delta_{ij}$. Thus, formula \eqref{derivadadelett} gives the responses for the free system simply as
\be \label{nonisotropicexpressions}
\chi_{ij}^0=\frac 1{k_BT} \left\langle d_i |d_j\right\rangle .
\ee
Therefore, we can write the generalized susceptibility for the interacting case \eqref{derivadadelett} as
\be
\chi_{ij}= \sum_{k} \chi_{ik}^0 \Delta_{kj} .
\ee
By writing explicitly the form of the $\Delta_{ij}$ coefficients we get
\be
\chi_{ij}= \sum_{k} \chi_{ik}^0 \left[\delta_{kj} +\sum_l U_{kl}\langle d_l|d_j\rangle\right]^{-1}.
\label{chi}
\ee
(where $[M_{ij}]^{-1}$ entails for the $ij$ element of the inverse matrix). Now comparing with the isotropic results \eqref{isotropicexpressions}, we propose the definition of the generalized, matrix-valued, Landau parameters
\be \label{landauP}
\boxed{~
F_{ij}=\sum_l U_{il}\langle d_l|d_j\rangle.
\phantom{\!\!\!\!\!\frac{\frac{\frac{1^1}2}2}2}
}
\ee

To check whether these parameters play a similar role in Pomeranchuk instabilities, we have to refer to our previously published results in ref. [\onlinecite{nos5}]. There, we have shown that a Pomeranchuk instability at zero temperature appears whenever the matrix
\be
M_{ij}^0= \delta_{ij}+F_{ij}
\ee
has a negative eigenvalue. By comparing this with \eqref{lachota} we see that indeed our definition for anisotropic Landau parameters (eq \eqref{landauP}) is playing a similar role, but in a matrix-like way. 

For most of the interactions studied in the literature one can choose a basis of functions $\{{d}_i^{\alpha}(\vk)\}_{i\in\mathbb{N}}$ such that the matrix $F_{ij}$ is diagonal,
\be
F_{ij}=F_{(i)}\delta_{ij}.
\label{diagonalizatechango}
\ee
In this new basis the stability condition can be written as
\be
1+F_{(i)}>0,
\label{estabilizategato}
\ee
which is now exactly equal to \eqref{lachota}. Moreover, the susceptibility \eqref{chi} reads
\be
\chi_{ij}=\chi_{ij}^0(1+F_{(i)})^{-1},
\ee
to be compared to \eqref{isotropicexpressions}. Thus, the definition for the generalized Landau parameters \eqref{landauP}, when the interaction function allows a diagonalization of the matrix, leads to the same expressions as the one for the the isotropic case.
\subsection{Perturbations on the interaction strengths}

We can also consider the case of perturbations to the interaction coefficients $U_{ij}$. In other words the interaction strengths $U_{ij}$ may play the role of control parameters. In this case the derivatives would be
\be
\frac{d\epsilon_o^\alpha(\vk)}{dU_{kl}}=0
\quad \quad
\frac{d U_{ij}}{dU_{kl}}=\frac 1{k_BT}  \left(
\delta_{ik}\delta_{jl}+\delta_{il}\delta_{jk}
\right),
\ee
implying that the response in the order parameters $\eta_j$ is given by
\be 
\label{derivadadeletatar}
\chi_{jkl}=\frac{d\eta_j}{dU_{kl}}
-\frac1{k_BT}\sum_r
\left(
\eta_k
\langle d_{{l}} | d_r \rangle
+
\eta_l
\langle d_{{k}} | d_r \rangle
\right) 
\Delta_{rj}.  
\ee
The expression measures the response of the system when a specific channel of the interaction is modified. An experimental realization of this process is possible in cold atom physics, where the interactions between atoms can be changed by means of Feshbach resonances \cite{reviewcoldatoms}. 
\subsection{Perturbations in both the dispersion relation and interaction strengths: thermal response and effective mass} 

Now we examine applications of our formula \eqref{derivadadeletata} to situations in which both terms in the parenthesis are non-vanishing. 

The first situation is that of the thermal response of an order parameter $\eta_j$. In this case the control parameter is the temperature, meaning $a_i=T$. We will denote this by writing $a_T=T$. The derivatives we should consider now are
\be
\frac{d\tilde\epsilon_o^\alpha(\vk)}{dT}=-\frac 1{kT^2}\epsilon_o^\alpha(\vk)
\quad\quad
\frac{d\tilde U_{ij}}{dT}=-\frac 1{kT^2}U_{ij}.
\ee
We get for the thermal response
\be
\chi_{jT}=-\frac 1{k_BT^2}
\sum_{r} 
\left(\left\langle\epsilon_o| d_{ r} \right\rangle 
-\sum_{ {k}{l}}
\eta_{{k}} U_{{k}{l}} \langle d_{{l}} | d_{ r} \rangle \right)
\Delta_{{r}{j}}.  
\ee
The last expression can be rewritten as
\be
\chi_{jT}=-\frac 1{k_BT^2}
\sum_{ r} 
\left\langle\epsilon| d_{ r} \right\rangle 
\Delta_{{r}{j}}.  
\ee

As a particular example, to calculate the specific heat we should take the derivative of the internal energy $U = \langle H \rangle$ with respect to the temperature. Then the internal energy will play the role of our order parameter $\eta_\epsilon=U$. In the mean field approximation it reads
\be
U \equiv \eta_\epsilon= \sum_{\vk,\alpha} \,F[\tilde\epsilon^\alpha(\vk)]
\epsilon^\alpha(\vk).
\ee
This allows us to identify 
\be \label{gammasusc}
c_V=\chi_{\epsilon T}=-\frac 1{k_BT^2}
\sum_{ r} 
\left\langle\epsilon| d_{ r} \right\rangle 
\Delta_{{r}\epsilon}.  
\ee

A straightforward calculation also gives the thermal response of the magnetization $\eta_a$ and the number of particles $\eta_s$ as
\be
\chi_{aT}=
-\frac1{k_BT^2}
\sum_{ r} 
\left\langle\epsilon| d_{ r} \right\rangle 
\Delta_{{r}a}  
\ee
and
\be
\chi_{sT}=
-\frac1{k_BT^2}
\sum_{ r} 
\left\langle\epsilon| d_{ r} \right\rangle 
\Delta_{{r}s}.  
\ee

~

We can also calculate an effective mass tensor as follows. Lets assume that we create an excitation with momentum $\bf P$ on the ground state of the system described by the Hamiltonian \eqref{hamil}. The total momentum of the system is now $\bf P$, and it can be boosted back to the rest frame by redefining the sum variable as $\vk=\underline{\vk}+{\bf P}$ and the number operators as $n^\alpha(\underline{\vk}+{\bf P})\equiv  {\underline{n}}^\alpha(\underline{\vk})$. Then expanding the dispersion relation in powers of ${\bf P}$ we get the same mean field Hamiltonian as before in terms of $\underline{\vk}$ and $\underline{n}^\alpha(\underline{\vk})$, but with a perturbation of the form (we omit the underline in $\vk$ in what follows)
\be
\Delta \epsilon_o^\alpha(\vk)=
{\bf P}\cdot  \nabla_{\bf k}\epsilon_o^\alpha(\vk), 
\ee
\small
\be
\Delta f^{\alpha\beta}(\vk,\vk') =
\sum_{ij}\!U_{ij} {\bf P}\cdot
\left(
\nabla_{\vk} d_i^\alpha(\vk)d_j^\beta(\vk')
\!+\!
d_i^\alpha(\vk)\nabla_{\vk'} d_j^\beta(\vk')
\right).
\ee
\normalsize
Calling $d_{{\sf a}i}^\alpha(\vk)=\partial_{\sf a}d_i^\alpha(\vk)$, we see that new couplings $U_{ij}{\bf P}_{\sf a}$ arise in the interaction, multiplying the functions $d_{{\sf a}i}^\alpha(\vk)$ and $d_{j}^\alpha(\vk)$. We can identify
\be
\frac{d\tilde\epsilon_o^\alpha}{dP_{\sf a}}=\frac1{k_BT}\epsilon^\alpha_{o{\sf a}}(\vk)
\quad\ \
\frac{d\tilde U_{ij}}{dP_{\sf a}}=
\frac1{k_BT}
(\delta_{i,{\sf a}k}U_{kj}+\delta_{j,{\sf a}k}U_{ki}),
\ee
An inverse effective mass tensor can then be defined as the response of the velocity
\be
{\bf V} = \sum_{\vk,\alpha}\, \nabla_{\bf k}\epsilon^\alpha(\vk) 
\ee
to such perturbation. We obtain
\be
[m^{-1}_*]^{\sf ab}
=
\frac1{k_BT}\sum_{ r} 
\left(\left\langle  \epsilon_{{\sf a}}|d_{ r} \right\rangle 
+
\sum_{ {k}{l}}
\eta_{{\sf a}{k}}  U_{{k}{l}} \langle d_{{l}} | d_{ r} \rangle 
\right)
\Delta_{{r}{\epsilon_{\sf b}}}.  
\ee
\section{Application examples}
\label{sec:interactions}

\begin{figure}
   \includegraphics[width=.23\textwidth]{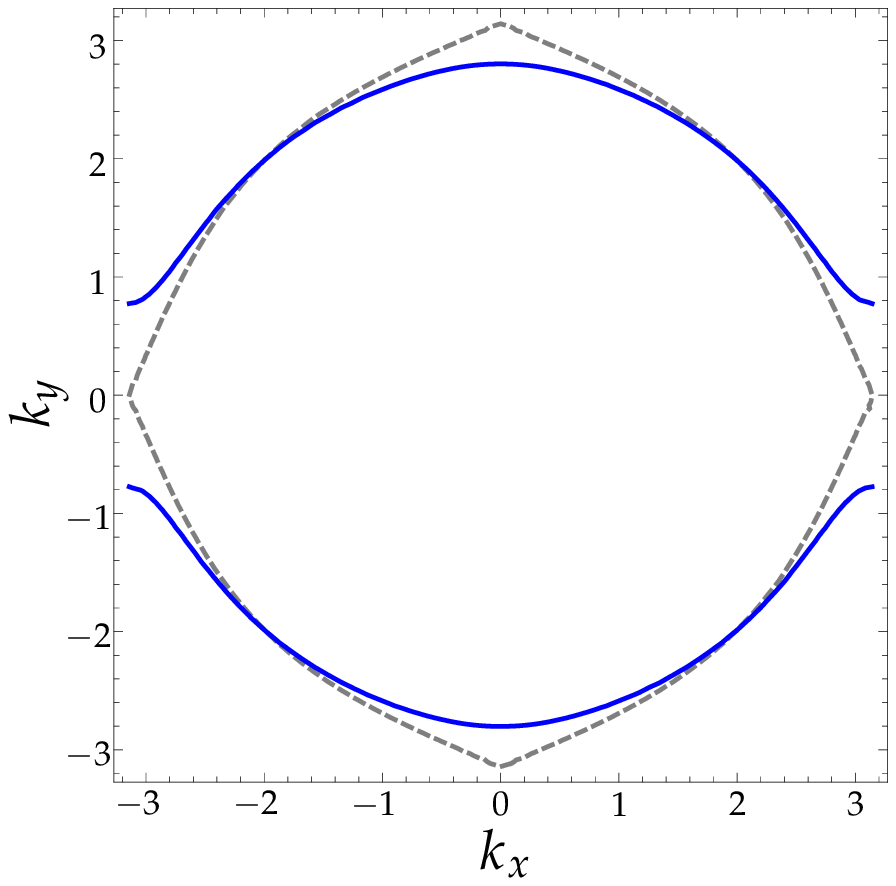}
   \includegraphics[width=.23\textwidth]{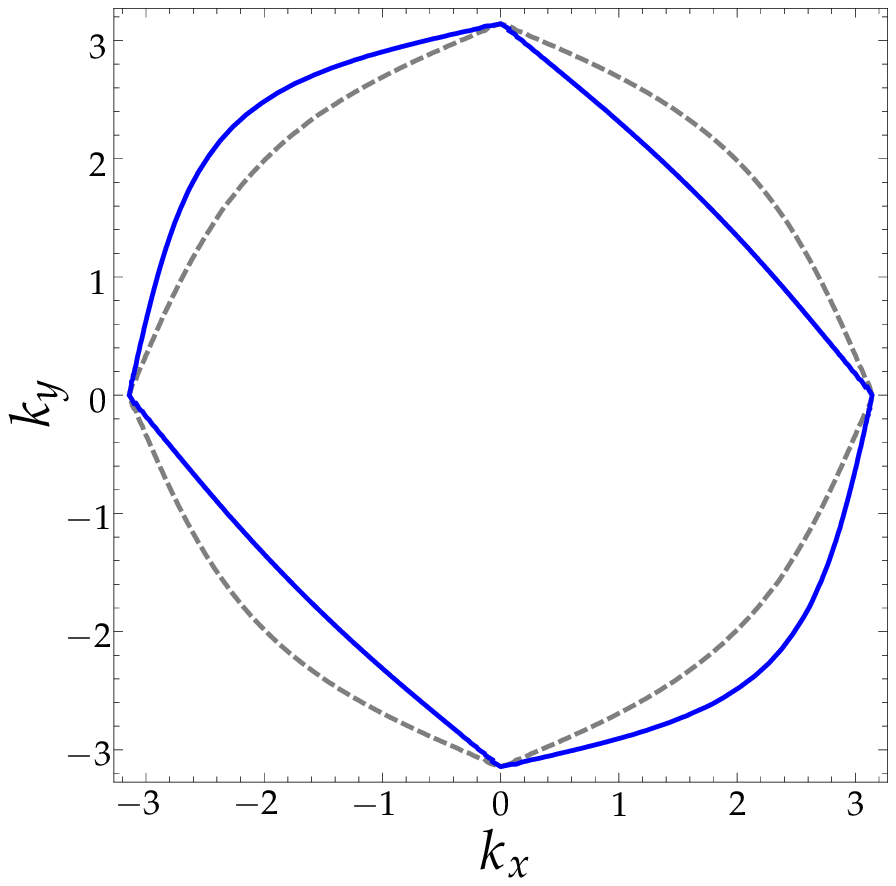}
\caption{(Color online) Distorted Fermi surfaces obtained considering the interaction with $d_{x^2-y^2}$ (left) and $d_{xy}$ (right), both are compared with the Fermi surface obtained for the  non interacting case (first and second neighbors hopping) at Van Hove filling (gray dashed lines). We see that the distortions introduced in the second case do not avoid the Van Hove singularities.}
\label{fig:fsdeformadas}
 \end{figure}

\begin{figure} 
   \includegraphics[width=.23\textwidth]{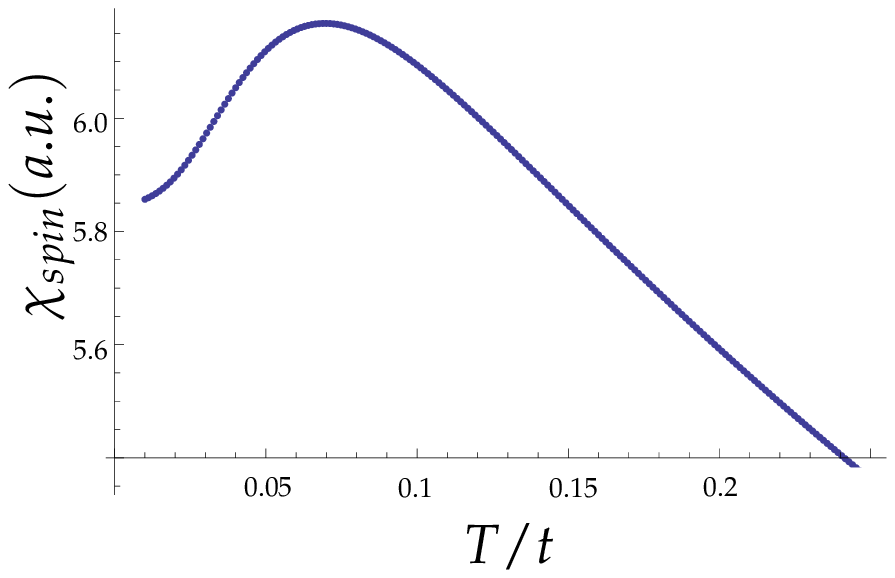}
   \includegraphics[width=.23\textwidth]{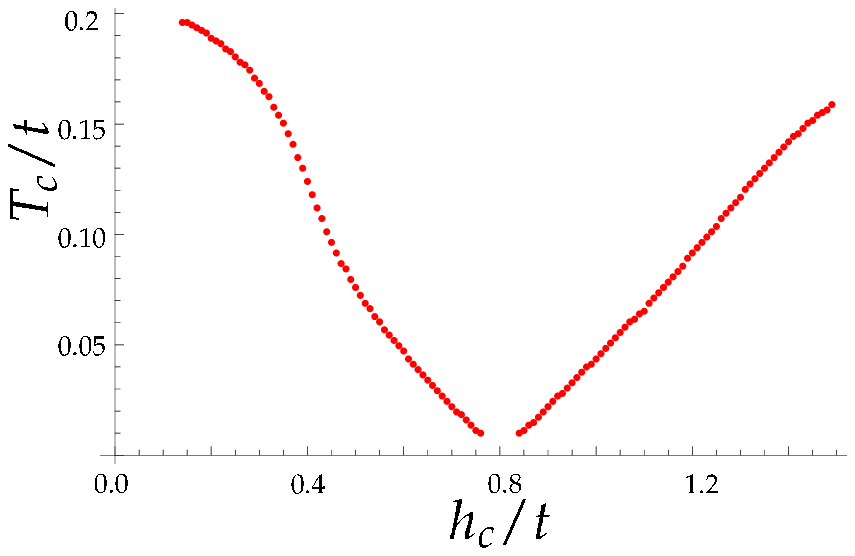}
\caption{(Color online) Non interacting system ($U=0$). Left panel: it is shown the susceptibility as a function of magnetic field for $h=0.5t$. Right panel: we constructed an estimation for the crossover curve by tracking the maximum of the susceptibility for each point of the phase diagram.}
\label{fig:yamaselibre}
 \end{figure}
In this section we illustrate the previous results by computing the response functions in the case where the interaction function has only one term in the expansion \eqref{interac}. We consider a spin-independent interaction which reads 
\be \label{interasimple}
 f(\mathbf{k},\mathbf{k'}) = -U d(\mathbf{k}) d(\mathbf{k'}),
\ee

In this case, we can explicitly calculate the matrix elements $\Delta_{ij}$, introduced in eq. \eqref{Delta}, as

\be
\Delta=
\left( \begin{array}{ccc}
1 & 0 & 0 \\
0 & 1 & 0 \\
\frac{-U \langle d|d_s \rangle}{1+ U \langle d|d \rangle}  & \frac{-U \langle d|d_a \rangle}{1+ U \langle d|d \rangle}  & \frac{1}{1+ U \langle d|d \rangle}  \end{array} \right).
\ee
Replacing in expression \eqref{spinsusc} we have the spin susceptibility. It reads

\be \label{yamasespinsusc}
\chi_{spin}= \frac{1}{k_B T} \left( \langle d_a|d_a \rangle   - \frac{ U \langle d_a|d \rangle^2}{1+U \langle d|d \rangle} \right).
\ee
Similarly, from \eqref{chargesusc} and \eqref{gammasusc}, we calculate the charge susceptibility and the specific heat. They read
\be \label{yamasechargesusc}
\chi_{charge}= \frac{1}{k_B T} \left( \langle d_s|d_s \rangle   - \frac{ U \langle d_s|d \rangle^2}{1+U \langle d|d \rangle} \right)
\ee
and
\be \label{yamasegamma}
c_V= \frac{1}{k_B T^2} \left(-\langle \epsilon|\epsilon \rangle + \frac{ U \langle \epsilon|d \rangle^2}{1+ U \langle d|d \rangle}   \right),
\ee
with the previously introduced notation (eq. \eqref{notation}). 

The matrix containing the Landau parameters, which is given by equation \eqref{landauP}, is now
\be
F=
\left( \begin{array}{ccc}
0 & 0 & 0 \\
0 & 0 & 0 \\
U \langle d|d_s \rangle  &U \langle d|d_0 \rangle  & U \langle d|d \rangle  \end{array} \right).
\ee

~

In the next paragraphs we calculate this responses for particular interaction functions. We study the cases of d-wave form factors $d_{x^2-y^2}$ and $d_{xy}$. That means $d(\mathbf{k})= \cos{k_x} - \cos{k_y}$ in the former case and $d(\mathbf{k})= \sin{k_x}\sin{k_y}$ in the latter. 
We take the bare dispersion relation corresponding to a $2$-dimensional square lattice with first and second neighbors hopping\cite{yamase2010}
\be \label{reldisperyamase}
\varepsilon_o(\vk)= - 2 \left(t( \cos{k_x} +\cos{k_y})+ 2 t' \cos{k_x} \cos{k_y}\right)\,,
\ee
For both d-wave form factors we study the systems at (or near) the van-Hove filling, where Pomeranchuk instabilities \cite{Pomeranchuk} are enhanced and Fermi liquid phases with different shapes of Fermi surface may emerge \cite{nos5}. In Fig.\ref{fig:fsdeformadas}, the deformed Fermi surfaces, obtained via the renormalized dispersion relation \eqref{reldisperrenor}, are depicted and compared with the unperturbed ones. We see that the van-Hove singularities (points $(0,\pm\pi)$ or $(\pm\pi,0)$ on the first Brillouin zone) are avoided when the symmetry of the deformation is the one of the $d_{x^2-y^2}$ form factor. On the other hand, in the case of $d_{xy}$ interaction the Fermi surface deformation do not change the instability scenario.

~

\subsection{Interaction with $d_{x^2-y^2}$ form factor}

We study a model, described by the Hamiltonian \eqref{hamil}, characterized by an interaction function with single-term interaction with the $d_{x^2-y^2}$ form factor 
\be
d(\mathbf{k})= \cos{k_x} - \cos{k_y},
\ee
and the dispersion relation of a $2$-dimensional square lattice with first and second neighbors hopping, eq. \eqref{reldisperyamase}. It has been recently proposed to describe the phenomenology of $Sr_3Ru_2O_7$ at low temperatures \cite{yamase2010,grigera2011,grigera2012}. Following reference [\onlinecite{yamase2010}], we take for the second neighbors hopping $t'=0.35$, the interaction strength is $U=1$, and the chemical potential which controls the filling $\mu=1$, all measured in units of the first neighbors hopping $t$. We also add a coupling with a uniform external magnetic field $h$.
It is worth noticing that, when considering this interaction, the expressions \eqref{yamasespinsusc}-\eqref{yamasegamma} are the same that have been previously calculated by Yamase and co-workers\cite{yamase2010,yamaserestob}.

Let us first consider the non-interacting case ({\em i.e.} $U=0$). In Fig.\ref{fig:yamaselibre}a we show the susceptibility for $h=0.5t$. There we see that there is a critical temperature $T_{c}$, located around $T/t=0.65$, where the response changes its behavior. Collecting the values of $T_{c}$ for different values of magnetic field, obtained as described above, we can estimate the crossover curve (shown in Fig.\ref{fig:yamaselibre}b).

\begin{figure}
   \includegraphics[width=.48\textwidth]{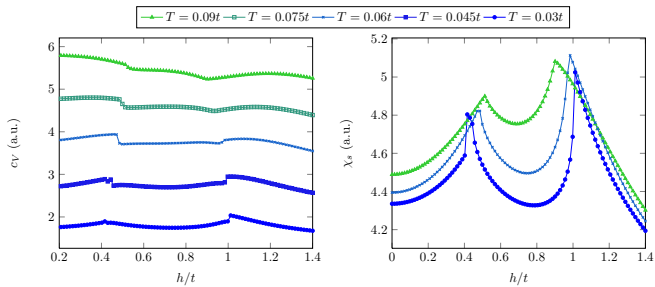}
\caption{(Color online) When considering the interactions ($U=t$) we see jumps in the specific heat (left) and peaks in the susceptibilities (right), corresponding with the phase transitions.}
\label{fig:yamasinteracting}
 \end{figure}

\begin{figure}
   \includegraphics[width=.4\textwidth]{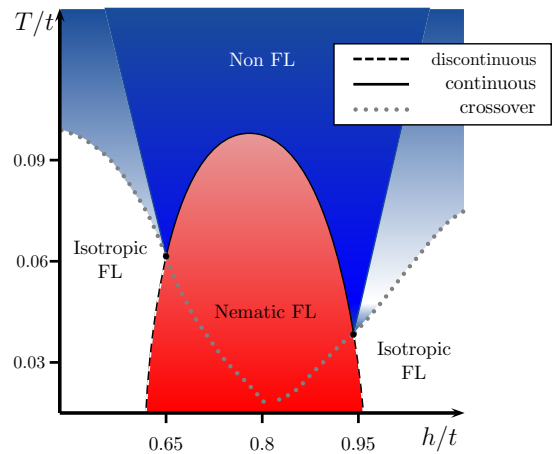}   
\caption{(Color online) Schematic phase diagram which captures the low temperature behavior of $Sr_3Ru_2O_7$.}
 \label{fig:diagramarutenato}
\end{figure}

When we turn on the interactions ($U=t$) we obtain the specific heat and spin susceptibility as a function of the magnetic field, shown in Fig.\ref{fig:yamasinteracting}. The specific heat curves corresponding to lower temperatures show discontinuities while entering or leaving the symmetry-broken phase. On the other hand, for higher temperatures, the curves are smoothed indicating a continuous phase transition. Also the susceptibilities show peaks for the same magnetic field values. All these results are consistent with the phase diagram exhibited by the $Sr_3Ru_2O_7$ \cite{yamase2010,yamaserestob,nos5}, schematized in Fig.\ref{fig:diagramarutenato}.

\subsection{Interaction with $d_{xy}$ form factor}

Now we consider the same dispersion relation (eq. \eqref{reldisperyamase}) and a single-term interaction with the $d_{xy}$ form factor, 
\be
d(\mathbf{k})= \sin{k_x}\sin{k_y}.
\ee
To obtain a non-trivial solution for the mean-field free energy ({\em i.e.} a region in the space of parameters with non-vanishing order parameter), we need to consider a strong interaction, with a coupling constant of about 10 times the first neighbors hopping ($U/t\simeq10$). 

\begin{figure}[bp] 
  \includegraphics[width=.23\textwidth]{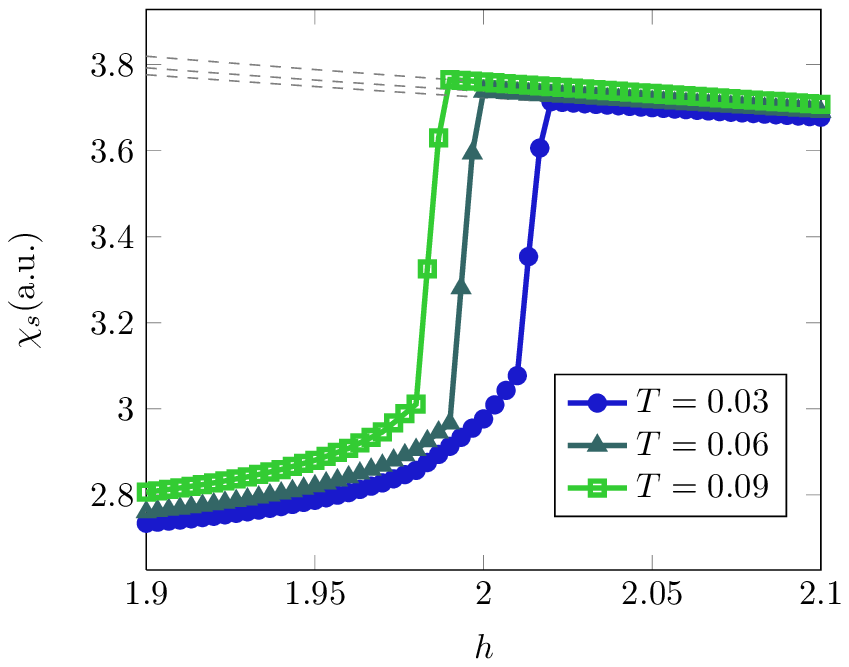}
  \includegraphics[width=.23\textwidth]{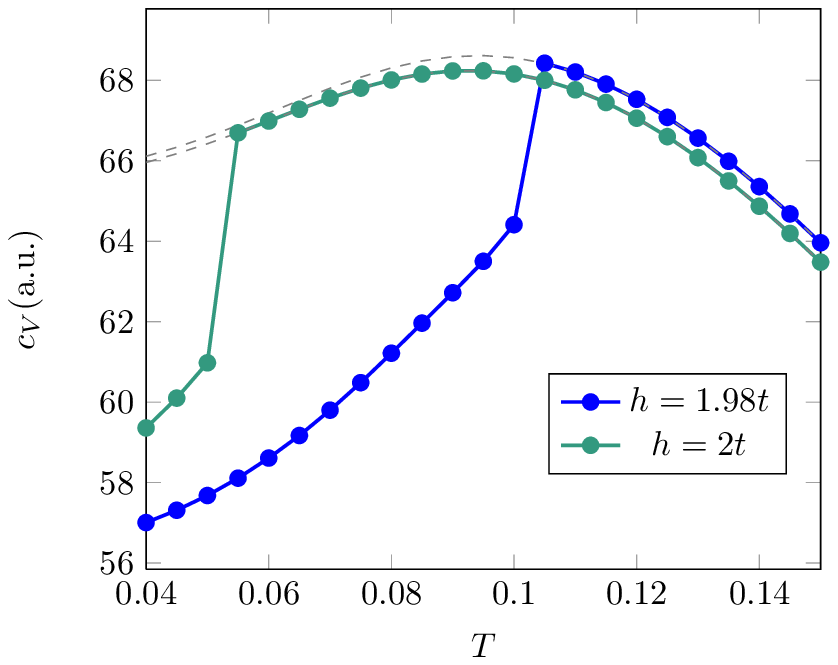} \\
\caption{ (Color online) Left: magnetic susceptibility curves as a function of magnetic field for different temperatures. Right: specific heat coefficient $h=1.98t$ and $h=2t$. In both panels non-interacting responses are shown for comparison (gray,dashed).}
\label{fig:dxyresults}
\end{figure}

Besides, when we analyze the stability of that Fermi liquid phase, following Pomeranchuk criteria \cite{nos4, nos5}, we find that it is stable only for lower temperatures and magnetic fields around $h=2t$. In this region of the space of parameters, we use the response relations to compute the magnetic susceptibility and the specific heat \eqref{yamasegamma}, for different temperatures. The results are shown in Fig.\ref{fig:dxyresults}.

\section{Conclusions}
\label{sec:conclusions}

In the present paper we developed a simple way to treat Fermi liquids when the Fermi surface is not rotationally invariant. Using a generic decomposition of the interaction function as in \eqref{interac} we were able to obtain simple expressions for the linear response functions. These expressions generalize those for the isotropic Fermi liquid to the case where the Fermi surface lacks continuous rotational symmetry. In particular we analyzed the responses obtained under changes in external control parameters such as the magnetic field, the density or the temperature and we obtain the spin and charge susceptibilities, as well as a specific heat and the effective mass tensor. In addition, we also took a particular channel of the interaction as a control parameter and we were able to obtain a simple expression for the corresponding response function.  

By comparing our results with the standard expressions for the isotropic Fermi liquid, we propose a natural definition for the generalized version of the Landau parameters, which have in general a matrix structure. With the help of such Landau parameters, one can relate the responses to those corresponding to the non interacting limit.

For illustration, we applied our results to cases where the interaction function has only one term in the expansion and the dispersion relation corresponds to that of a lattice system (and thus leads to an anisotropic Fermi surface). For the case of an interaction consisting on a $d_{x^2-y^2}$ form factor in a square lattice with first and second neighbors hopping, our results are compatible with previous calculations \cite{yamase2010,yamaserestob}. 
When considered a $d_{xy}$ form factor, we found jumps in the specific heat at the discontinuous phase transitions.

The reported results could be useful in the characterization of Fermi liquid phases on lattices, by allowing simple calculations of static susceptibilities even when the Fermi surface lacks continuous rotational invariance.
Also, they provide a simple test to phenomenological models, by checking whether they present divergences on the thermodynamic responses.

\section*{Acknowledgments}

This work was partially supported by PICT ANPCyT (grant No 1724), PIP CONICET (grants No 0396 and 0747) and UNLP (projects X529, X648 and X659).

\appendix
\section{Mean Field approximation}
\label{sec:meanfield}
In this section we explicit the mean field approximation we have performed\cite{Quintanilla3, meanfield} on a generic Hamiltonian $H$ describing the Fermi liquid with four fermion interaction. We start with
\be
\begin{split}
\hat{H} &= \sum_{\vk,\alpha}  \epsilon_o^\alpha(\vk) \hat{c}^\dagger_{\alpha,\vk} \hat{c}_{\alpha, \vk}  \\
&+ \sum_{\vk,\vk',\textbf{q}} \sum_{\alpha,\beta,\gamma, \delta} f^{\alpha\beta\gamma \delta}(\vk,\vk',\textbf{q})  \hat{c}^\dagger_{\alpha, \vk+\textbf{q}} \hat{c}^\dagger_{\gamma, \vk'-\textbf{q}} \hat{c}_{\delta, \vk'} \hat{c}_{\beta, \vk}, 
\end{split}
\ee
where $\hat{c}^\dagger_{\alpha, \vk}$ y $\hat{c}_{\alpha, \vk}$ are the fermionic creation and annihilation operators, and 
$\epsilon_o^\alpha(\vk)$ is the dispersion relation. We approximate it by a diagonal mean field Hamiltonian $\hat{H}_{0}$
\be
\hat{H}_{0}=\sum_{\alpha,\vk} \epsilon^\alpha(\vk) \hat{c}^\dagger_{\alpha, \vk}\hat{c}_{\alpha, \vk}.
\ee
Within our approximation the mean values are computed as
\be
 \left\langle ...\right\rangle_0 = \frac{1}{Z_{0}}Tr\left( e^{-\beta \hat{H}_{0}} ... \right),
\ee
where $Z_{0}=Tr (e^{-\beta \hat{H}_{0}} )$, $\beta= 1/{k_BT}$ and the mean-field free energy is given by
\be
\begin{split}
 F_{0}
 &=-\frac{1}{\beta}\sum_{\vk,\alpha} \log\left(1+e^{-\beta \epsilon^\alpha(\vk)}\right).
\end{split}
\ee

With this at hand, the original expression for the free energy can be written as
\be
F= -\frac{1}{\beta}\ln(Tr\{e^{-\beta \hat{H}}\}) =
-\frac{1}{\beta}\ln(
Tr\{ e^{-\beta  (\hat{H}- H_{0})}   e^{ -\beta  H_{0}}  \}),
\ee
and we can expand the exponential, yielding
\be
F =-\frac{1}{\beta} \log Tr \left( e^{-\beta \hat{H}_0} - (\hat{H}-\hat{H}_0)e^{-\beta \hat{H}_0} +\cdots \right).
\label{tatata}
\ee
The approximate expression for the free energy is finally given by
\begin{equation}\label{mfquintanilla}
F \simeq F_0 + \langle \hat{H}-\hat{H}_{0} \rangle_0. 
\end{equation}
Where we have taken the mean field approximation, that is assuming that the dots in (\ref{tatata}) can be discarded. By minimizing this expression we find the renormalized dispersion relation $\epsilon^\alpha(k)$. To perform the calculation lets begin by analyzing the term
\be
\begin{split}
\langle \hat{H}-\hat{H}_{0}  \rangle_0= \sum_{\alpha,\vk}( \epsilon_o^\alpha(\vk)&- \epsilon^\alpha(\vk) )   \langle  \hat{c}^\dagger_{\alpha, \vk}\hat{c}_{\alpha, \vk} \rangle_0  \\
+ \frac{1}{2} \sum_{\alpha,\beta,\gamma,\delta} \sum_{\vk,\vk',\textbf{q}} &f^{\alpha\beta\gamma\delta}(\vk,\vk',\textbf{q})
\\ \times & \langle \hat{c}^\dagger_{\alpha, \vk+\textbf{q}} \hat{c}^\dagger_{\gamma, \vk'-\textbf{q}}\hat{c}_{\delta, \vk'}c_{\beta, \vk} \rangle_0. 
\end{split}
\ee

By using Wick's theorem and the low temperature limit where $\langle \hat{c}^\dagger_{1} \hat{c}_{2} \rangle_0 = \delta_{12} n_{1}$, we have

\ba
\langle \hat{H}-\hat{H}_{0}  \rangle_0&=& \sum_{\alpha,\vk}( \epsilon_o^\alpha(\vk)- \epsilon^\alpha(\vk) )  n_{\alpha,\vk}  \nonumber \\
+&\frac{1}{2}& \biggl( \sum_{\alpha,\beta} \sum_{\vk,\vk'} f^{\alpha\alpha\beta\beta}(\vk,\vk',\textbf{0}) n_{\alpha,\vk} n_{\beta,\vk'} \\
&-&\sum_{\alpha\beta} \sum_{\vk,\vk'}  f^{\alpha\beta\beta\alpha}(\vk,\vk',\vk'-\vk) n_{\alpha,\vk'} n_{\alpha,\vk} \biggr). \nonumber
\ea
where $n_{\alpha,\vk} = F[\epsilon_\alpha(\vk)/k_BT]$.
In the forward scattering limit $\textbf{q}\to \textbf{0}$, we only consider the contributions of the first sum containing the interaction function $ f^{\alpha\alpha\beta\beta}(\vk,\vk',\textbf{0})\equiv f^{\alpha\beta}(\vk,\vk')$. 
We get
\be \begin{split}
\langle \hat{H}-\hat{H}_{0}  \rangle_0 =& \sum_{\alpha,\vk}(\epsilon_o^\alpha(\vk) - \epsilon^\alpha(\vk) )  n_{\alpha,\vk} \\
&+\frac{1}{2} \sum_{\alpha,\beta} \sum_{\vk,\vk'} f^{\alpha\beta}(\vk,\vk') n_{\alpha,\vk} n_{\beta,\vk'}.  
\end{split}
\ee
Using this expression the approximation for the free energy reads
\be \begin{split}
F =& \sum_{\vk,\alpha}(\epsilon_o^\alpha(\vk) - \epsilon^\alpha(\vk) )  n_{\alpha,\vk}
\\
&+\frac{1}{2} \sum_{\alpha,\beta} \sum_{\vk,\vk'} f^{\alpha\beta}(\vk,\vk')  n_{\alpha,\vk} n_{\beta,\vk'} \\
&- \frac{1}{\beta}\sum_{\vk,\alpha} \log[1+e^{-\beta \epsilon^\alpha(\vk)}], \nonumber
\end{split} \ee
where $ n_{\alpha,k}=(1+e^{-\beta \epsilon^\alpha(\vk)})^{-1} $. We find $\epsilon^\alpha(\vk) $ from the condition ${\delta F}/{\delta\epsilon^\alpha(\vk)}=0$.
Thus,
\be
\epsilon^\alpha(\vk)= \epsilon_o^\alpha(\vk)+ \frac{1}{2} \sum_{\beta, \vk'} \biggl(f^{\alpha \beta}(\vk,\vk') + f^{\beta \alpha}(\vk',\vk) \biggr) n_{\beta,\vk'} .
\ee
The symmetry of the interaction function leads to
\be
\epsilon^\alpha(\vk)= \epsilon_o^\alpha(\vk) + \sum_{\beta,\vk'} f^{\alpha \beta}(\vk,\vk') n_{\beta,\vk'} .
\ee
By using the decomposition of the interaction function, eq. \eqref{interac}, we obtain the renormalized dispersion relation of eqs. \eqref{reldisperrenor}-\eqref{orderparameter}
\ba
\epsilon^\alpha(\vk)&=&\epsilon_o^\alpha(\vk)
-\sum_{ij}^N U_{ij}\eta_{i} d_{j}^\alpha(\vk), \\ 
\eta_{i}&=&\sum_{\alpha,\vk} \, d_{i}^\alpha(\vk) F[\epsilon^\alpha(\vk)].
\ea

~
 
We can now write our approximate free energy, from \eqref{mfquintanilla}, as
\be \label{ourFenergy}
F = \frac{1}{2} \sum_{ij}^N  U_{ij}\eta_{i} \eta_{j} 
- \frac{1}{\beta}\sum_{\vk,\alpha} \log[1+e^{-\beta \epsilon^\alpha(\vk)}].
\ee
From this expression we can obtain the order parameters just by taking derivatives. For example, by adding to the hamiltonian the perturbation (\ref{zeeman}) and then taking the derivative with respect to the magnetic field, we can get the magnetization, as
\be
M=\frac{dF}{dh} =  \sum_{ij}^N  U_{ij} \frac{ d\eta_{i}   }{d h} \eta_{j} 
- \sum_{\vk,\alpha} \frac{d \epsilon^\alpha(\vk)}{dh}n_{\alpha,\vk},
\ee
where we have used the symmetry $U_{ij}=U_{ji}$. By using the expression for the renormalized dispersion relation we obtain 
\be
M  =- \sum_{\vk,\alpha} \frac{d \epsilon_o^\alpha(\vk)}{dh} n_{\alpha,\vk}=- \sum_{\vk,\alpha} \alpha n_{\alpha,\vk}.
\ee

\newpage

\bibliographystyle{apsrev4-1}
\bibliography{biblio}

\end{document}